**Olesya Mryglod[1,2], Yurij Holovatch[2,3]**
[1] Institute of Computer Sciences and Information Technologies, Lviv Polytechnic National University, 12 S. Bandery Str., Lviv, 79013, UKRAINE
[2] Institute for Condensed Matter Physics of the NAS of Ukraine, 1 Svientsitskii Str., 79011 Lviv, UKRAINE
[3] Institut für Theoretische Physik, Johannes Kepler Universität Linz, 69 Altenbergerstr., 4040 Linz, AUSTRIA

# Human activity as the decision-based queueing process: statistical data analysis of waiting times in scientific journals

© Olesya Mryglod, Yurij Holovatch

*Ми розглядаємо процес редакційної обробки статей в наукових журналах як процес людської активності, що базується на основі прийняття рішень. Використовуючи класичні підходи теорії масового обслуговування, фізики критичних явищ та статистичні методи аналізу даних, ми вивчаємо функціональну форму ймовірнісного розподілу випадкових величин, що описують динаміку людської поведінки. Додатковою метою є обґрунтування наукометричного застосування отриманих результатів.*

*We consider the editorial processing of papers in scientific journals as a human activity process based on the decision making. A functional form of the probability distributions of random variables describing a human dynamics is studied using classical approaches of mass service systems theory, physics of critical phenomena and statistical methods of data analysis. Our additional goal is to corroborate the scientometrical application of the results obtained. Key words – data analysis, statistics, mass service systems, human activity, scientometrics*

## Introduction

Nowadays new possibilities for quantitative studying of human activity processes appear. The modern computer technologies allow us to collect and to process large volumes of statistical data [1–5]. A human dynamics can be considered as the set of consecutive actions in time. Moreover, the tasks that an individual has to do, form some queue which exist in memory, in notebook, in reminder, etc. At every moment of time an individual makes a decision to execute some task from the queue based on some perceived priority [2]. And indeed, the situation, when individual has some number of tasks and executes them dependently on their subjective priorities, – is natural.

The analysis of time statistics of human activity patterns can be useful for different optimization and control tasks in spheres of mass service, communication, information technologies, resource distribution, etc. Besides, this approach can give us new possibility to understand human behaviour and to get its additional quantitative measure. The classical models of human dynamics, used from risk assessment to communications, assume that human actions are randomly distributed in time and thus well approximated by Poisson processes. But the results of recent investigations made evident the non-Poisson statistics for timing of many human dynamics processes: the long periods of inactivity separated by bursts of intensive activity [2,5]. The power-law (or close to it) functional form of probability distributions:

$$P(t) \sim t^{-\alpha} \qquad (1)$$

founded in human activity processes can be the reason for drawing some analogies with critical phenomena in physics [7]. On the other hand, the processes with execution of tasks in queue are the typical objects of study in the classical mass serving theory [6]. Thus, different methods and approaches can be used to analyse human activity processes.

New researches of human activity processes are based on the possibility to operate with numerous statistical data about different human activity processes such as browsing the Internet, data downloading, electronic communication, initiating financial transactions etc [1–5]. In such studies two random variables are often considered: the time interval between two consecutive actions by individual (called the interevent time $t_{int}$) and the so called waiting time $t_w$, the time a task is waiting for an execution. In the case of priority-based (decision-based) queuing process the timing of the tasks will be heavy tailed, because the most tasks being rapidly executed, whereas a few experience very long waiting times [2]. It is interesting to note, that the exponent that governs the power-law dependence differs for different queuing processes and its value can be dependent on existence or absence of restrictions on queue length [5]. It is important, that power-law-like distributions (1) appear only in critical and supercritical regimes of service, when its traffic intensity $\rho = \lambda/\mu$ is $\geq 1$ ($\lambda$ is the arrival rate and $\mu$ is the execution rate of tasks, respectively). In other words, the probability distribution functions of human activity processes are close to power law only when the queue of tasks exists [5].

## The problem statement

We consider the process of editorial revision in scientific journals as an example of human activity processes [7,8]. In this kind of mass service system the input flow consists of submitted papers forming the queue. A standard procedure after paper submission can include following steps: (i) work of referees, (ii) corrections if necessary, (iii) acceptance by an Editorial Board. On each of the above stages the paper may be rejected. However, typically the information concerning the rejected papers is not publicly available. Therefore, we consider the random variable $t_w$ defined as a number of days between the dates of the paper final acceptance $t_a$ and the paper receiving $t_r$:

$$t_w = t_a - t_r. \qquad (2)$$

All stages of the editorial editing of scientific papers are considered together as the one process (Fig. 1). Though more than one actor takes part in it, we consider that every part of this work is controlled by an Editorial Board. That's why we can treat this process as one of the main characteristics of the Editorial Board work.

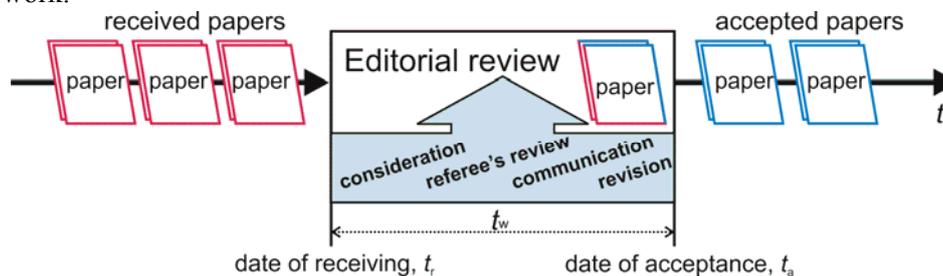

*Fig. 1. Schematic picture of editorial processing of papers in scientific journals.*

Our goal is to determine the functional form of probability distributions $P(t_w)$ based on the statistical data analysis performed for a few scientific journals. Another task is to determine if possible any typical form of $P(t_w)$ for normally working Editorial Board.

## Statistical data

Three scientific editions of the international Elsevier Publishing House with different Editorial Boards were chosen in our study: "Physica A: Statistical Mechanics and its Applications", "Physica B: Condensed Matter" [7,8] and "Information Systems". The publicly accessible information on official web-cite (*http://www.elsevier.com*) about dates of receiving and acceptance for scientific papers has been used to calculate waiting times $t_w$. Some formal parameters of databases created for these journals are shown in Table 1.

*Table 1*
**Characteristics of analysed databases**

|  | "Physica A" | "Physica B" | "Information Systems" |
|---|---|---|---|
| Number of records | 2667 | 2692 | 740 |
| Maximal value of $t_w$ | 1629 days | 1087 days | 2260 days |
| The minimal value of $t_w$ for all journals equals 1 day. | | | |

## Data analysis

First of all, we can calculate and compare the average time intervals $\langle t_w \rangle$ which papers has to wait for acceptance in the chosen scientific journals. Moreover, the dynamics of $\langle t_w \rangle$ through the years can be observed. For example, we found the average waiting times in "Physica A" and "Physica B" show a tendency to decrease, while $\langle t_w \rangle$ for the "Information Systems" journal increases with time (Fig. 2).

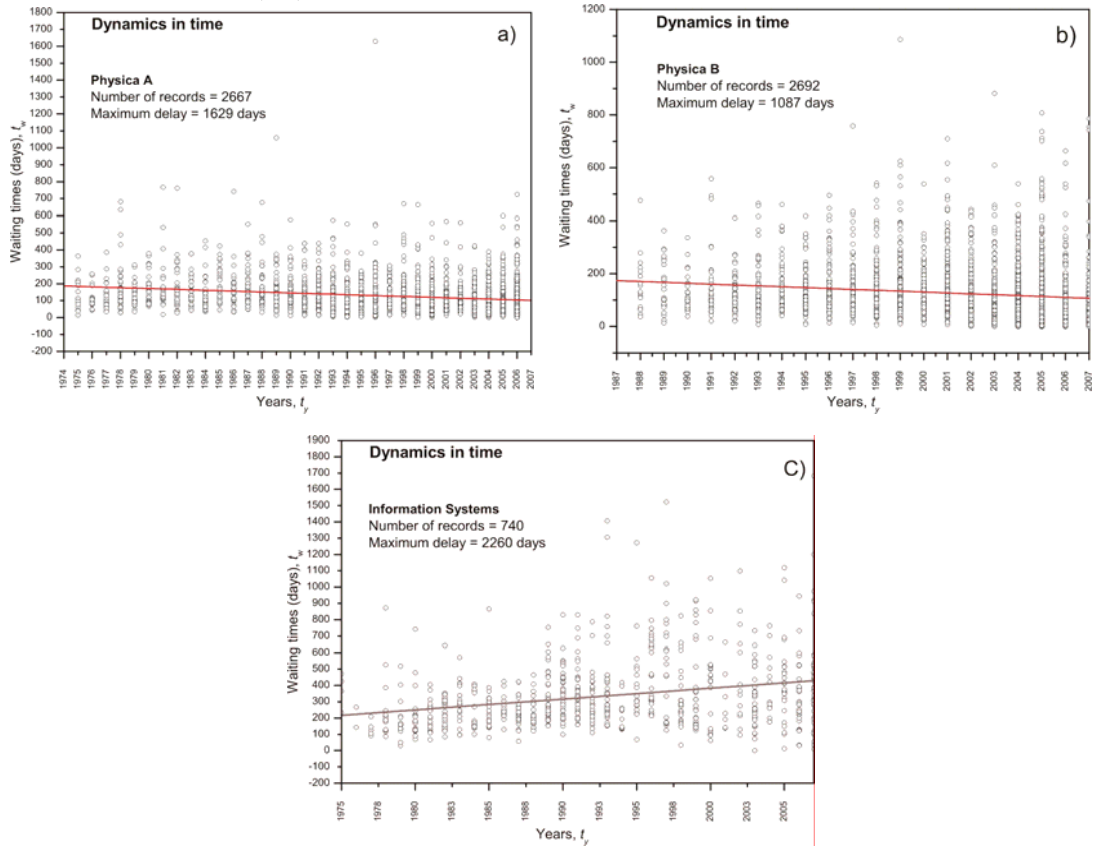

*Fig. 2. Time dynamics of average editorial processing time in editions: (a) "Physica A", (b) "Physica A" and (c) "Information Systems". Every point denotes waiting time of one paper in the respective year.*

Further steps of our research were connected with a study of functional form of the probability distributions $P(t_w)$. To avoid strong data fluctuations we have built the cumulative probability distributions $P_>(t_w)$ of waiting time $t_w$ [7,8], namely:

$$P_>(t_w) = \int_{t_w}^{t_w^{max}} P(t_w) \, dt_w, \qquad t_w^{min} \leq t_w \leq t_w^{max}, \qquad (3)$$

where $t_w^{min}$ and $t_w^{max}$ are the minimal and maximal waiting times, respectively. $P_>(t_w)$ shows the number of papers with waiting time greater than $t_w$. The cumulative distributions $P_>(t_w)$ of waiting time were built for all observed journals in both log-log and log-normal scales. The results of linear approximations performed on several intervals of cumulative distribution $P_>(t_w)$ for "Information Systems" journal in log-log and log-normal scales are shown in Fig. 3. We can see the good possibilities for linear approximations in both scales and this situation is analogous for "Physica A" and "Physica B" journals [7]. Thus the analysis of cumulative distributions $P_>(t_w)$ is useless to find the functional form of probability distribution of waiting time $t_w$.

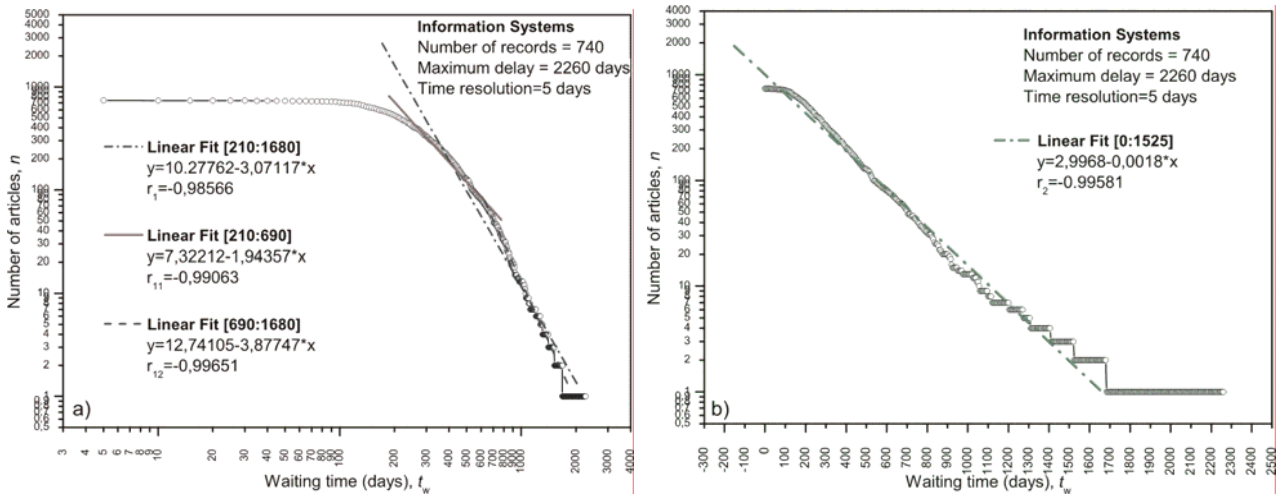

*Fig. 3. The cumulative distributions $P_>(t_w)$ of waiting time $t_w$ built for "Information Systems" journal in (a) log-log and (b) log-normal scales.*

Further, for better accuracy the probability histograms $P(t_w)$ were constructed. We have verified two main hypotheses about form of probability distributions which describe human activity processes: (i) log-normal distribution [4]

$$P(t_w) = P_0 + \frac{A}{\sqrt{2\pi}\,\varpi\,t_w}\,e^{-\frac{\left[\ln\left(\frac{t_w}{t_c}\right)\right]^2}{2\varpi^2}}, \qquad t_c, \varpi > 0, \qquad (4)$$

where $\ln t_c$ and $\varpi$ are the mean and standard deviations of the $\ln(t_w)$, $P_0, A$ are fitting constants; and (ii) power-law distribution with exponential cutoff [5] for exponent values $\alpha = \{1; 3/2\}$

$$P(t_w) = A t_w^{-\alpha}\,e^{-\frac{t_w}{t_0}}, \qquad t_0 > 0, \qquad (5)$$

where $t_0$ is characteristic of waiting time which depends on traffic intensity, $A$ is a constant.

Our results obtained for all three journals are shown in Fig. 4. Using fitting procedure we found optimal parameters for both distributions (4) and (5). To compare the accuracy of approximations by different functions we have calculated the values of chi-square correlation coefficients: $R = \chi^2/\text{DoF}$, where $\chi^2 = \sum (t_w - t_{\text{theor}})^2 / t_{\text{theor}}$ and DoF denotes statistical degrees of freedom. The resulting values of correlation coefficients $R$ and the parameters of approximations are shown in Table 2.

*Table 2*
**The values of correlation coefficients and parameters of approximations**

| Theoretical law | „Physica A", $t_w \in [85;770]$ | „Physica B", $t_w \in [85;810]$ | „Information Systems", $t_w \in [200;1060]$ |
|---|---|---|---|
| Log-normal (4) | $R \approx 5.54 \cdot 10^{-4}$ if $P_0 \approx 6.8 \cdot 10^{-4}$, $\varpi \approx 0.63$, $A \approx 4.44$, $t_c \approx 106.5$ | $R \approx 5.46 \cdot 10^{-4}$ if $P_0 \approx 7 \cdot 10^{-4}$, $\varpi \approx 0.65$, $A \approx 4.3$, $t_c \approx 102.8$ | $R \approx 30.24 \cdot 10^{-4}$ if $P_0 \approx 0$, $\varpi \approx 0.53$, $A \approx 7.37$, $t_c \approx 278$ |
| Power-law with exponential cutoff (5) and $\alpha = 1$ | $R \approx 4.65 \cdot 10^{-4}$ if, $A \approx 4.74$, $t_0 \approx 204.6$ | $R \approx 4.33 \cdot 10^{-4}$ if $A \approx 4.39$, $t_0 \approx 210.2$ | $R \approx 17.07 \cdot 10^{-4}$ if $A \approx 9$, $t_0 \approx 498$ |
| Power-law with exponential cutoff (5) and $\alpha = 3/2$ | $R \approx 5.97 \cdot 10^{-4}$ if $A \approx 36.3$, $t_0 \approx 469.9$ | $R \approx 5.05 \cdot 10^{-4}$ if $A \approx 33.6$, $t_0 \approx 500.6$ | $R \approx 17.76 \cdot 10^{-4}$ if $A \approx 103.8$, $t_0 \approx 1319.8$ |

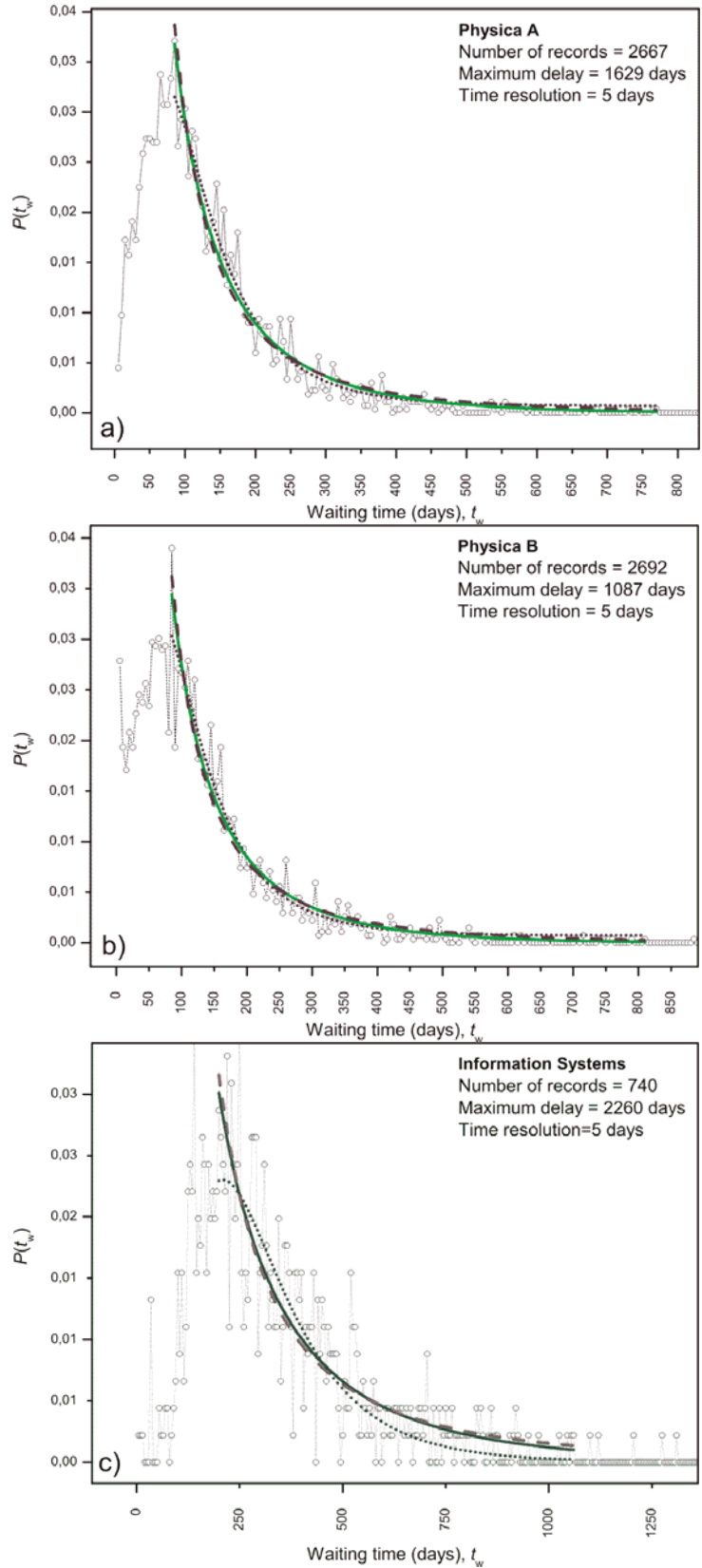

*Fig. 3. The probability histograms $P(t_w)$ of waiting time $t_w$ built for (a) "Physica A", (b) "Physica B" and (c) "Information Systems" journals. The dotted dark lines are the approximations by log-normal (4), the light solid and dark dashed lines – by power-law with exponential cutoff with $\alpha = 1$ and $\alpha = 3/2$ (5), respectively.*

**Conclusions**

We have found that both log-normal and power-law function with exponential cutoff and exponent $\alpha = 1$ can be the probable functions of distributions $P(t_w)$. The same conclusions have been done for all three journals. In fact, both log-normal and power-law functions predict exactly the same leading behavior $t^{-1}$, differing only in the functional form of the exponential correction [9]. Thus, we can state that the studied case of human activity is in accordance with other researches in this field [1–5].

The observed data fluctuations can be explained by relatively small statistics but such situation is usual for the majority of scientific journals. Thus, we consider the obtained form of probability distributions $P(t_w)$ as the typical one that can be used for scientometrical analysis of editions. The length of time period which papers waiting for publication is important characteristic of Editorial Board's work. The publication delay effects on journal rankings according to the impact factor as well as on personal citation rating of authors [10].